\documentclass[aps]{revtex4}
\usepackage{epsfig,graphics,amssymb,amsmath,subeqnarray,color}

\def \u{\mathbf{u}}\def\n{\mathbf{n}}\def\E{\mathbf{E}}
\def\x{\mathbf{x}}\def \y{\mathbf{y}}\def \z{\mathbf{z}}
\def \e{\mathbf{e}}\def \r{\mathbf{r}}\def \0{\mathbf{0}}
\def \n{\mathbf{n}}

\def\0{\mathbf{0}}

\def\ord1{^{(1)}}\def\ords{^{(2)}}
\let\originaleqref=\eqref
\renewcommand{\eqref}{Eq.~\originaleqref}

\begin{document}

\title{Phase-separation models for swimming enhancement in complex fluids}
\author{Yi Man}
\author{Eric Lauga}
\email{e.lauga@damtp.cam.ac.uk}
\affiliation{
Department of Applied Mathematics and Theoretical Physics, 
University of Cambridge, CB3 0WA, United Kingdom.}
\date{\today}
\begin{abstract}

Swimming cells often have to self-propel through fluids displaying non-Newtonian rheology. While past theoretical work seems to indicate that stresses arising from complex fluids  should systematically hinder low-Reynolds number locomotion, experimental observations suggest that locomotion enhancement is possible. In this paper we propose  a physical mechanism for locomotion enhancement of microscopic swimmers  in a complex fluid. It is based on the fact that micro-structured fluids will generically phase-separate near surfaces, leading to the presence of low-viscosity layers which  promote slip and decrease viscous friction near the surface of the swimmer. We use two models to address the consequence of this phase separation: a nonzero  apparent slip length for the fluid and then an explicit modeling of the change of viscosity in a thin layer near the swimmer. Considering two {\color{black} canonical setups} for low-Reynolds number locomotion, namely the waving locomotion of a two-dimensional sheet and that of a three-dimensional filament, we show that phase-separation systematically increases the locomotion speeds, possibly by orders of magnitude. We close by confronting our predictions with  recent experimental results.  

\end{abstract}
\maketitle
\section{Introduction}

Over the past few decades, problems on  life at low Reynolds number have received significant attention,  from both the physics and biological communities \cite{Taylor51,lighthill75,Purcell1977,Winet1977,braybook}. For the most part,  theoretical and experimental studies have focused on cell locomotion in Newtonian fluids, with an emphasis on the interplay between biological actuation and whole-cell response, for example the relationship between cell geometry, waving actuation,  and the resulting swimming kinematics \cite{Winet1977,Powers2009}.

Studies on cell motility in fluids typically  focus on one of four types of  cells -- bacteria \cite{bergbook}, spermatozoa \cite{gaffney11}, ciliates \cite{blake74}, and single-celled  planktonic cells \cite{stocker} -- while  recent effort  considered  larger, multi-cellular organisms \cite{goldstein_annurev}. In all cases, the  biological environments that the microorganism  encounter may be rheologically complex.  For example, \emph{Helicobacter pylori}, a bacterium that causes inflammation in the stomach, swims through  gastric mucus to protect itself from the acidic environment \cite{hpPNAS09}. Mammalian spermatozoa have to progress   through  highly-elastic  cervical mucus, an important phase in  reproductive process \cite{sperm78}. 

Extending our understanding of cell locomotion in Newtonian fluids to complex, gel-like or viscoelastic environments is a nontrivial task. One quintessential question, whether non-Newtonian stresses in a  complex fluid help increase or decrease the swimming speed of the cell, remains in many ways an open problem.

For example, bacteria such as \emph{Leptospira} and \emph{Escherichia coli},  swim more rapidly in gel-like unbranched polymer solutions than in Newtonian fluids \cite{BergTurner79}. In contrast, for the nematode \emph{Caenorhabditis elegans}  undergoing undulatory swimming, the speed was observed to decrease in a slightly shear-thinning polymeric fluid with strong elastic stresses \cite{Shen11}. Similar  disparities are observed experimentally for bio-inspired synthetic swimmers. Force-free rotating  helices show a transition from hindered to enhanced swimming  in 
constant-viscosity Boger fluids \cite{Liu11} while a cylindrical version of Taylor's swimming sheet  displays both increase and decrease as a function of the  rheology of the fluid \cite{Powers13}.  In contrast, externally-actuated flexible-tail swimmers show a systematic increase of locomotion speeds in viscoelastic fluids
\cite{zenit13}.

Various numerical and theoretical studies have also addressed this problem, focusing on viscoelastic fluids following 
 Oldroyd-B rheology. Small-amplitude asymptotic studies for waving swimmers with fixed shapes predicted a systematic decrease of swimming velocity \cite{Lauga07,Fu07,Fu09}. Subsequent numerical work    for finite waving sheets with large tail amplitude showed that an increase was possible for order one Deborah numbers \cite{Teran10}. Numerical simulations following the helical experiments in Ref.~\cite{Liu11} confirmed the transition from slow small-amplitude swimming to fast large-amplitude locomotion  \cite{Saverio13}. Integral theorems for small-amplitude motion showed that the superposition  of multiple waves could also lead to a enhancement transition for a range of Deborah numbers \cite{lauga14}. 

We thus see that theoretical, computational, and experimental studies showing both increases and decreases have been put forward, and the challenge is now to rigorously untangle the various physical (and sometimes, biological) effects. In particular, while we  now understand how viscoelastic stresses are able to decrease swimming speeds, physical mechanisms leading to locomotion enhancement are less clear. Recently, the flexibility of the swimmer  in response to complex stresses was shown to  allow for an increase in the swimming speed \cite{guy14,riley14}. In this paper, we propose a different {\color{black} physical origin} for the observed swimming enhancement. Instead of focusing on the new non-Newtonian stresses in the fluid, we address one of the consequences of having a structured fluid,  namely the fact that it is expected to phase-separate near the body of the swimmer, leading to the well-known phenomenon of  apparent slip.

For a variety of complex  fluids with a microstructure dispersed in a solvent, in particular  polymeric fluids and suspensions, the presence of a boundary leads to static phase separation at equilibrium: the  concentration of the solute drops  near the wall which is covered instead by a thin solvent layer  \cite{Barnes95}. In the case of rigid  suspensions, purely excluded-volume interactions lead to solvent-rich regions near the surface, and the effect is larger for Brownian particles for which the  presence of a wall breaks the geometrical isotropy \cite{Barnes95}. In the case of polymers, random coils would be distorted if too close to the wall, and thus they are driven by entropy away from the boundary.

\begin{figure}[t]
\centering
\includegraphics[width=5in]{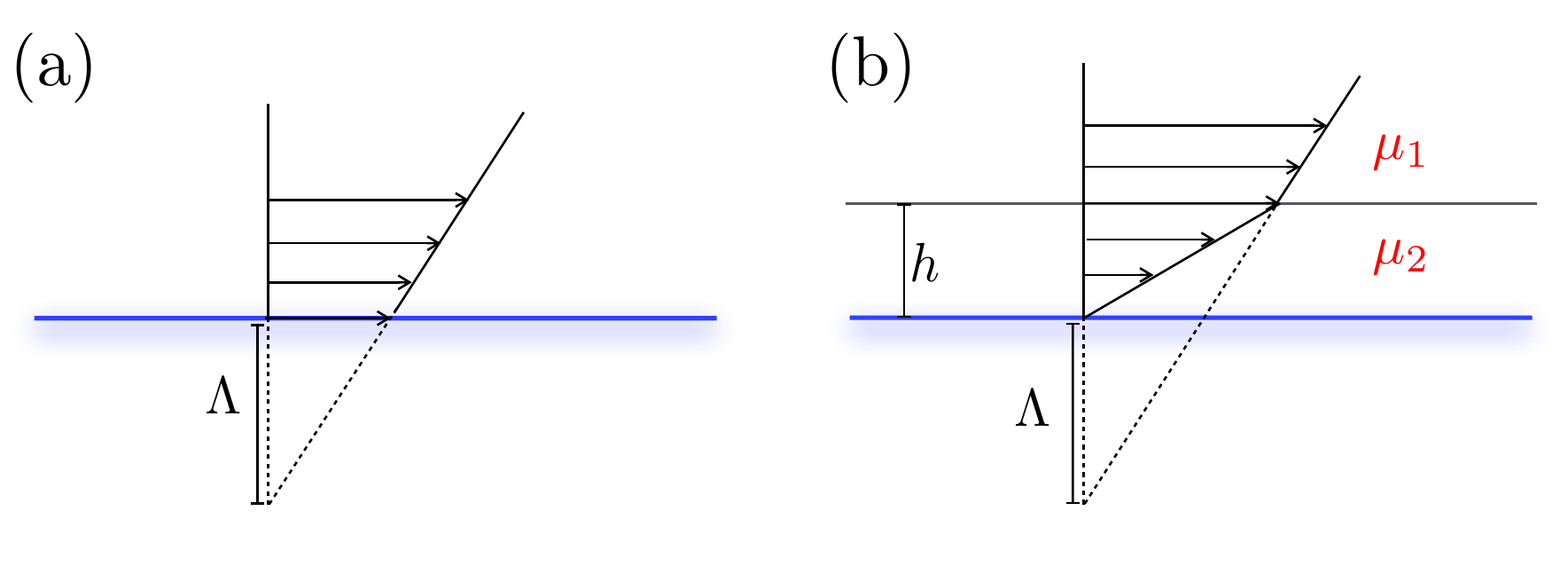}
\caption{{\color{black}{(Color online)}} The two models of apparent slip due to phase separation considered in this paper; 
(a): a single-phase continuum fluid with a finite apparent  slip length $\Lambda$; 
 (b): a two-fluid domain with  viscosity $\mu_1$ in the bulk   and a thin low-viscosity layer of thickness $h$ and  viscosity $\mu_2 < \mu_1$.}\label{shear2fluids}
\end{figure}

In all cases, the solvent-rich fluid near the surface has a viscosity much smaller than that of the bulk fluid. As seen in  many situations, in particular flows in capillary tubes and in  porous media \cite{Cohen85}, this difference in viscosity leads to  apparent slip when a flow is set up, which is best illustrated in the case of a  shear flow (Fig.~\ref{shear2fluids}b): If shear  is imposed in a fluid with a thin-viscosity layer, the difference in viscosities will lead to a difference in shear rates, and as a result the flow in the high-viscosity bulk will not extrapolate to zero on the solid surface, but below it, indicating an overall decrease of stresses acting on the surface. The fictitious distance below the surface where the fluid velocity in the top fluid goes to zero is the (positive) apparent slip length. Microscopically, the no-slip condition is of course not violated, but given that the typical thickness of the solvent layer is much smaller than the other, macroscopic length scales in the problem of interest, the no-slip boundary condition appears not to hold for the bulk fluid.

There are two classical  ways to theoretically model apparent slip in complex fluids. The first model is to simply replace the no-slip boundary condition on the surface by one which the tangential velocity is allowed to slip. Experimentally-measured slip length has been shown to depend, sometimes in a complex manner, on the shear stress at the wall \cite{sliplength1,Kalyon91,Kalyon93,Cohen85}. The assumption usually done is to adopt Navier's slip length model \cite{navier} and assume that the slip velocity at the wall is linearly proportional to the wall shear rate, with a proportionality constant with unit of length, called the slip length, and which we will denote $\Lambda$ in this work \cite{Sochi11}.  As noted above, the slip length measures the 
(fictitious)  distance below the boundary  where the velocity would extrapolate to zero, and it is zero in the case of a no-slip boundary  (Fig.~\ref{shear2fluids}a). A second procedure to model phase separation is to explicitly assume the presence of two fluid layers. The top layer, semi-infinite, has bulk viscosity $\mu_1$, while the bottom layer near the surface has a finite thickness $h$ and lower viscosity $\mu_2 < \mu_1$ (Fig.~\ref{shear2fluids}b). For a shear flow in this unidirectional setup, the velocity in the thin layer satisfies the no-slip boundary condition while that the flow in  the bulk fluid extrapolates to zero at  the 
equivalent apparent slip length $\Lambda=(\mu_1/\mu_2-1)h$.

In this paper we consider these two different physical models of phase separation and investigate their consequences on waving locomotion.  In 
\S\ref{sec:slip} we first examine the model with a finite apparent slip length, and apply it to two {\color{black} canonical setups} for low-Reynolds number locomotion, namely the small-amplitude swimming of a flexible sheet  \cite{Taylor51} and that of a flexible filament   \cite{Taylor52}. In  
\S\ref{sec:two-fluid}  we then consider the same two {\color{black} setups} in the situation where the phase-separated fluid is modeled as a two-fluid layer. In all cases we are able to derive  the swimming speed for each {\color{black} swimmer} analytically and we compare it to the case for the homogenous  Newtonian fluid with a no-slip boundary condition. We demonstrate that the phase separation leads to a systematic enhancement of the swimming speed, and suggest that this might play a role in the recently-measured swimming enhancement at low-Reynolds numbers.

\section{Swimming in a fluid with finite apparent slip length}
\label{sec:slip}
In the first section we assume that the phase separation in the fluid can be adequately captured by an effective slip length $\Lambda$ acting on a Newtonian fluid satisfying the Stokes equations
\begin{equation}\label{S}
\nabla p = \mu \nabla^2 \u,\quad \nabla\cdot\u = 0.
\end{equation}
On a  fluid-solid boundary $S$, the jump in normal velocity is zero by mass conservation while the jump in tangential velocity is proportional to the local shear rate. If the velocity in the fluid is denoted $\u$, these boundary conditions can be mathematically expressed as 
\begin{subeqnarray}\label{eq:bc}
\big[\n\cdot\u\big]\big|_{S} &=& 0,\\
\big [\n\times\u \big ]\big|_{S}&=& 
2\Lambda(\n\times(\E\cdot\n))\big|_{S}, 
\end{subeqnarray}
where $\big[...\big]$ is used to denote a jump, $\n$ is the normal to the boundary, $\E$ is the symmetric rate-of-strain tensor (i.e.~the  symmetric part   of the velocity gradient tensor), and $\Lambda$  the slip length. 

\subsection{Two dimensional waving sheet}
\label{sec:sheet-slip}
\begin{figure}[t]
\centering
\includegraphics[width=4in]{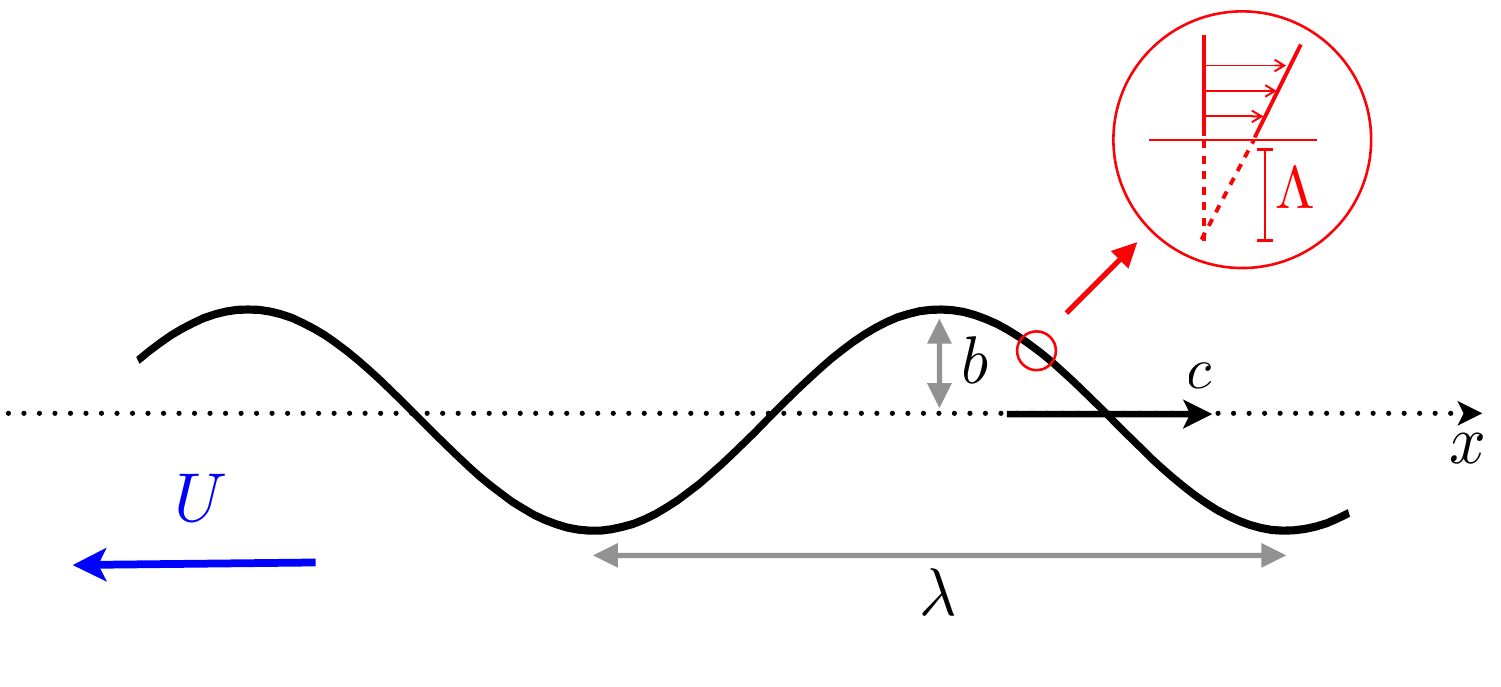}
\caption{{\color{black}{(Color online)}} Geometry of the two-dimensional waving sheet setup  in a fluid with finite slip length. Here $b$ denotes the small waving amplitude, $\lambda$ the wavelength, and $c$ the wave speed along the $x$ direction. The sheet is  assumed to swim with speed $U$ in the negative $x$ direction. The small circle zooms in on a portion of the swimmer surface and  illustrates the presence of a finite slip length $\Lambda$.}\label{geo_ws}
\end{figure}

We first consider a two-dimensional swimmer in the form of flexible sheet self-propelling in the fluid by passing waves of normal deformation. This is the classical setup originally proposed by Taylor \cite{Taylor51}, and the material points of the sheet, $(x_s,y_s)$, are assumed to vary in space and time as a simple traveling wave of deformation
\begin{equation}
y_s=b\sin k(x-ct),\quad x_s=x,
\end{equation}where $b$ is the wave amplitude, $k$ the wave number and $c$  the wave speed along the $x$ direction (see notation in Fig.~\ref{geo_ws}). We solve the problem assuming that  the amplitude is  small compared to the wavelength, and thus consider the asymptotic limit where $\epsilon=bk$ is a small dimensionless number. Nondimentionalizing the equations using $k^{-1}$ as relevant length and $\omega^{-1} \equiv (kc)^{-1} $ as intrinsic time scale, the wave deformation  becomes
\begin{equation}
\bar{y}_s=\epsilon\sin(\bar{x}-\bar{t}) \equiv  \epsilon\sin\xi,
\end{equation}
and for convenience we drop the ``bars'' in what follows. 

The  velocity of material points on   the sheet is thus given by 
\begin{equation}
\u_s=(0, -\epsilon \cos\xi).
\end{equation}
Applying  \eqref{eq:bc}, on the surface of the swimmer we have
\begin{subeqnarray}\label{eq:bc1_ws}
(\n\cdot\u)\big|_{(x_s, y_s)} &=& \n\cdot\u_s,\\
(\n\times\u)\big|_{(x_s, y_s)}&=& 2\bar{\Lambda}(\n\times(\E\cdot\n))\big|_{(x_s, y_s)}+\n\times\u_s,
\end{subeqnarray}
with a normal vector $\n$ explicitly given as  $(1+\epsilon^2\cos^2\xi)^{-\frac{1}{2}}[-\epsilon\cos\xi, 1]$, and where $\bar{\Lambda}\equiv k\Lambda$ is the nondimensionalized slip length. Here we keep the ``bar'' notation for the  slip length to ensure that the final result is formally dimensionless.

In order to solve for the Stokes equations in the fluid, \eqref{S}, we employ a streamfunction $\psi$ such that the velocity components of $\u = [u,v]$ are defined as 
 $\displaystyle u=\frac{\partial\psi}{\partial y}$ and $\displaystyle v=-\frac{\partial\psi}{\partial x}$. The boundary  condition,  \eqref{eq:bc1_ws}, then becomes
\begin{subeqnarray}\label{eq:bc2_ws}
\epsilon\cos\xi\frac{\partial\psi}{\partial y}(x, \epsilon\sin\xi)+\frac{\partial\psi}{\partial x}(x, \epsilon\sin\xi) &=& \epsilon\cos\xi,\\
\epsilon\cos\xi\frac{\partial\psi}{\partial x}(x, \epsilon\sin\xi)-\frac{\partial\psi}{\partial y} (x, \epsilon\sin\xi) &=&\bar{\Lambda} (1+\epsilon^2\cos^2\xi)^{-\frac{1}{2}}\bigg\{(1-\epsilon^2\cos^2\xi)\\&&\left[\frac{\partial^2\psi}{\partial x^2}(\xi, \epsilon\sin\xi)-\frac{\partial{\color{black}{^2}}\psi}{\partial y^2}(\xi, \epsilon\sin\xi)\right]\nonumber
\\ &&+4\epsilon\cos\xi\frac{\partial^2\psi}{\partial x\partial y}(\xi, \epsilon\sin\xi)\bigg\}+\epsilon^2\cos^2\xi.\nonumber
\end{subeqnarray}

The Stokes equation, Eq.~(\ref{S}), transforms into the biharmonic equation for $\psi$  \cite{leal}
\begin{equation}
\nabla^4 \psi = 0.
\end{equation}
 In order to obtain the  asymptotic solution for the swimming velocity, we  expand the streamfunction in powers of  $\epsilon$ by
\begin{equation}
\psi = \epsilon\psi^{(1)}+\epsilon^2\psi^{(2)}+\epsilon^3\psi^{(3)}+...
\end{equation}
Denoting the  velocity  of the swimming sheet as $-U\e_x$ in a quiescent fluid, we move in the swimming frame and thus the velocity at infinity is given by $\u(y\to \infty)= U\e_x$. 

Since $\psi$ satisfies the biharmonic equation, and is equal to $Uy$  at infinity, we construct the general solution as \cite{childress81}
\begin{subeqnarray}\label{eq:sf_g}
\psi^{(1)}&=&V_1^{(1)}+U^{(1)}y,\\
\psi^{(2)}&=&V_1^{(2)}+V_2^{(2)}+U^{(2)}y,
\end{subeqnarray}
where
\begin{equation}
V_n=(A_n+B_ny)e^{-ny}\sin n\xi+(C_n+D_ny)e^{-ny}\cos n\xi.
\end{equation}

 At  first order in $\epsilon$, Eq.~(\ref{eq:bc2_ws}) becomes
\begin{subeqnarray}
\frac{\partial\psi^{(1)}}{\partial x}(x,0)&=&\displaystyle \cos\xi,\\
-\frac{\partial\psi^{(1)}}{\partial y}(x,0) &=&\bar{\Lambda}  \left(\frac{\partial^2\psi^{(1)}}{\partial x^2}-\frac{\partial^2\psi^{(1)}}{\partial y^2}\right)\bigg|_{(x,0)}.
\end{subeqnarray}
Substituting Eq.~(\ref{eq:sf_g}) into the equation above, we  obtain $A_1^{(1)} = B_1^{(1)}=1$, $C_1^{(1)} = D_1^{(1)} = 0$ and $U^{(1)}=0$. 
The streamfunction at first order is
\begin{equation}
\psi^{(1)}=(1+y)e^{-y}\sin\xi,
\end{equation}
which is the same as the no-slip case. This can be rationalized by notating that the first-order shear rate is given by
\begin{equation}
\left(\frac{\partial^2\psi^{(1)}}{\partial x^2}-\frac{\partial^2\psi^{(1)}}{\partial y^2}\right)\bigg|_{(x,0)}=\left(-2ye^{-y}\sin\xi\right)\bigg|_{(x,0)}=0,
\end{equation}
which makes the problem equivalent to the no-slip case.

As no propulsion occurs at order $\epsilon$, one needs to carry the calculation to order two in order to obtain the leading-order swimming speed. At order $\epsilon^2$, the boundary conditions are 
\begin{subeqnarray}
\slabel{eq:ord21_ws}\left(\cos\xi\frac{\partial\psi^{(1)}}{\partial y}+\frac{\partial\psi^{(2)}}{\partial x}+\sin\xi\frac{\partial^2\psi^{(1)}}{\partial x\partial y}\right)\bigg|_{(x,0)}&=& 0,\\
\slabel{eq:ord22_ws}\left(\cos\xi\frac{\partial\psi^{(1)}}{\partial x}-\frac{\partial\psi^{(2)}}{\partial y}-\sin\xi\frac{\partial^2\psi^{(1)}}{\partial y^2}\right)\bigg|_{(x,0)}&=& \bar{\Lambda}  \left\{\left(\frac{\partial^2\psi^{(2)}}{\partial x^2}-\frac{\partial^2\psi^{(2)}}{\partial y^2}\right)\right.+\\&&\left.\sin\xi\left(\frac{\partial^3\psi^{(1)}}{\partial x^2\partial y}-\frac{\partial^3\psi^{(1)}}{{\color{black}{\partial y^3}} }\right)\right.\nonumber\\&&\left.+4\cos\xi\frac{\partial^2\psi^{(1)}}{\partial x\partial y}\right\}\bigg|_{(x,0)}+\frac{1}{2}+\frac{1}{2}\cos2\xi.\nonumber
\end{subeqnarray}
Substituting the expansions for the streamfunction into this condition, we  obtain
\begin{equation}
U^{(2)}=\frac{1}{2}+\bar{\Lambda}.
\end{equation}
Comparing this result with the no-slip case, and coming back to the dimensional variables we finally have
\begin{equation}
\frac{U^{(2)}}{U^{(2)}_{\rm no-slip}}=1+2 k \Lambda.
\end{equation}
Since the slip length is always positive, we obtain in this first situation that the  swimming speed is always enhanced by apparent slip.

\subsection{Three-dimensional waving filament}
\label{sec:3d:1}
\begin{figure}[t]
\centering
\includegraphics[width=2in]{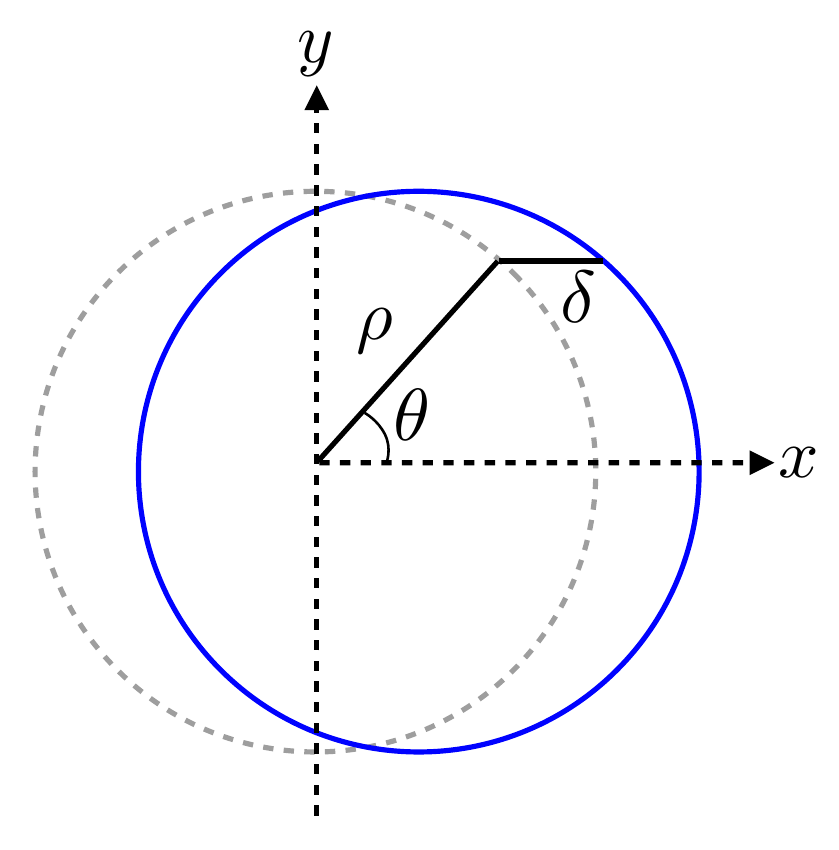}
\caption{{\color{black}{(Color online)}} Deformation with amplitude $\delta$ in the  circular cross-section  of a three-dimensional waving filament of radius $\rho$. The blue solid line represents the current position of the filament and the gray dashed line the average location of the cross section.}\label{geo_wc}
\end{figure}

We now apply the same apparent-slip model to the case of a three-dimensional waving filament, the geometry of which  is shown in Fig.~\ref{geo_wc} \cite{Taylor52}. We consider a cylindrical filament of radius $\rho$ deforming as a traveling wave in the $(x,z)$
plane where $z$ is along the filament axis and $x$ is perpendicular to it. We denote by $\delta$ the amplitude of the filament deformation in the $x$ direction.

The location of the material points on the surface of the filament, using  cartesian coordinate centered on the average location of the cylinder (see Fig.~\ref{geo_wc}), is given by
\begin{equation}
\r_s=(\delta+\rho\cos\theta)\x+\rho\sin\theta\ \y+z\ \z, \quad \delta=b\sin k(z+ct),
\end{equation}
where $\delta$ is the distribution of waving amplitudes along $z$ direction, $\rho$  the filament radius, $k$  the wave number, and $c$ the wave speed. Here again we assume that the  amplitude is  small compared to the
wavelength, and compute the result in the limit where  $\epsilon=bk$ is small.

Nondimentionalizing the equations similarly to the previous section, we have the surface of cylindrical filament  described as
\begin{equation}
\r_s=(\delta+\bar{\rho}\cos\theta)\x+\bar{\rho}\sin\theta\ \y+z\ \z, \quad \delta=\epsilon\sin(z+t)=\epsilon\sin s,
\end{equation}
and here again we keep the ``bar'' notation for the dimensionless radius, $\bar{\rho} = k \rho$.  Similarly to the two-dimensional case, the velocity on the boundary satisfies the conditions
\begin{subeqnarray}
(\n\cdot\u)\big|_{(\delta+\bar{\rho}\cos\theta,\bar{\rho}\sin\theta,z)}&=&\epsilon\cos s\cos\theta,\\
(\n\times\u)\big|_{(\delta+\bar{\rho}\cos\theta,\bar{\rho}\sin\theta,z)}&=&2\bar{\Lambda}\{\n\times(\E\cdot\n)\}\big|_{(\delta+\bar{\rho}\cos\theta,\bar{\rho}\sin\theta,z)}-\epsilon\sin\theta\cos s\ \z.
\end{subeqnarray}

Since the inextensibility condition contributes to the dynamics at  orders higher than two \cite{Taylor52},  the vector normal to the surface is   $\e_r$ at the order relevant for this calculation. We can then expand the velocity around the average position as
\begin{align}\label{eq:ex_ap}
\begin{split}
\u(\delta+\bar{\rho}\cos\theta, {\color{black}\bar{\rho}}\sin\theta, z)&\approx[\u+\x\cdot(\nabla\u)\epsilon\sin s]\big|_{r=\bar{\rho}}\\
&=\left\{u+\epsilon\sin s\left[\frac{\partial u}{\partial r}\cos\theta-\left(\frac{1}{r}\frac{\partial u}{\partial\theta}-\frac{v}{r}\right)\sin\theta\right]\right\}\bigg|_{r=\bar{\rho}}\e_{r}\\
&+\left\{v+\epsilon\sin s\left[\frac{\partial v}{\partial r}\cos\theta-\left(\frac{u}{r}+\frac{1}{r}\frac{\partial v}{\partial\theta}\right)\sin\theta\right]\right\}\bigg|_{r=\bar{\rho}}\e_{\theta}\\
&+\left\{w+\epsilon\sin s\left[\frac{\partial w}{\partial r}\cos\theta-\frac{1}{r}\frac{\partial w}{\partial\theta}\sin\theta\right]\right\}\bigg|_{r=\bar{\rho}}\e_{z}.
\end{split}
\end{align}

Expanding the velocity in the fluid,  $\u$, asymptotically in powers of  $\epsilon$, 
\begin{equation}
\u = \epsilon\u^{(1)}+\epsilon^2\u^{(2)}+\dots,
\end{equation}
and substituting the expansion and \eqref{eq:ex_ap} into the boundary condition, we obtain that at  first order,
\begin{subeqnarray}
u^{(1)}(\bar{\rho},\theta,z)&=&\cos s\cos\theta,\\
v^{(1)}(\bar{\rho},\theta,z)&=&\bar{\Lambda}\left(\frac{\partial v^{(1)}}{\partial r}-\frac{v^{(1)}}{r}+\frac{1}{r}\frac{\partial u^{(1)}}{\partial\theta}\right)\bigg|_{(r=\bar{\rho})}-\sin\theta\cos s,\\
w^{(1)}(\bar{\rho},\theta,z)&=&\bar{\Lambda}\left(\frac{\partial w^{(1)}}{\partial r}+\frac{\partial u^{(1)}}{\partial z}\right)\bigg|_{(r=\bar{\rho})}.
\end{subeqnarray}

Solving for the fluid velocity using separation of variables we get
\begin{equation}
u^{(1)}=u_q(r)\cos\theta\cos s,\quad
v^{(1)}=v_q(r)\sin\theta\cos s, \quad
\slabel{eq:bc2_wc_3}w^{(1)}=w_q(r)\cos\theta\sin s,
\end{equation}
while the first-order boundary conditions are simplified to
\begin{subeqnarray}\label{eq:bc2_wc}
u_q(\bar{\rho})&=&1,\\
v_q(\bar{\rho})&=&\bar{\Lambda}\left[v'_q(\bar{\rho})-\frac{v_q(\bar{\rho})}{\bar{\rho}}-\frac{1}{\bar{\rho}}\right]-1,\\
w_q(\bar{\rho})&=&\bar{\Lambda}\left[w'_q(\bar{\rho})-1\right].
\end{subeqnarray}

Similarly to the original problem treated by Taylor in the case of a no-slip filament \cite{Taylor52}, the radial dependence of the   velocity is given by a  combination of modified Bessel functions as
\begin{subeqnarray}\label{Bessels}
u_q(r)&=&BK_2(r)+CK_0(r)+ArK_1(r){\color{black}+EI_2(r)+FI_0(r)+DrI_1(r)},\\
v_q(r)&=&BK_2(r)-CK_0(r){\color{black}+EI_2(r)-FI_0(r)},\\
\slabel{Bessels3}w_q(r)&=&BK_1(r)+CK_1(r)+A[rK_0(r)-K_1(r)]{\color{black}-EI_1(r)-FI_1(r)}\nonumber\\
&&{\color{black}-D[rI_0(r)-I_1(r)]},
\end{subeqnarray}
{\color{black} with a minus sign in Eq.~\ref{Bessels3} coming from different  properties between solution of the  first and second kind}. {\color{black}For the boundary conditions at infinity, we have  $D=E=F=0$ and the other} three unknown constants $A, B, C$ can be obtained by plugging Eq.~(\ref{Bessels}) into  Eq.~(\ref{eq:bc2_wc}).  Writing $C$ as $C=C_{\rm nu}/C_{\rm de}$, we obtain the following lengthy (but analytical) expressions 
\begin{subeqnarray}
C_{\rm nu}&=&\bar{\rho} K_1(\bar{\rho})^2-2\bar{\rho} K_0(\bar{\rho})K_2(\bar{\rho})+2K_1(\bar{\rho})K_2(\bar{\rho})+\bar{\Lambda} \left[-\frac{\bar{\rho}}{2}K_0(\bar{\rho})K_1(\bar{\rho})\right.\\
&&\left.-K_0(\bar{\rho})K_2(\bar{\rho})+2K_1(\bar{\rho})^2+\left(\frac{4}{\bar{\rho}}-\frac{5}{2}\bar{\rho}\right)K_1(\bar{\rho})K_2(\bar{\rho})+K_2(\bar{\rho})^2\right]\nonumber\\
&&+\bar{\Lambda}^2\left[2K_0(\bar{\rho})K_1(\bar{\rho})-2\bar{\rho} K_1(\bar{\rho})^2+\frac{6}{\bar{\rho}}K_0(\bar{\rho})K_2(\bar{\rho})\right.\nonumber\\
&&\left.-6K_1(\bar{\rho})K_2(\bar{\rho})+\frac{2}{\bar{\rho} }K_2(\bar{\rho})^2\right],\nonumber\\
C_{\rm de}&=&-2\bar{\rho} K_0(\bar{\rho})^2K_2(\bar{\rho})+2K_0(\bar{\rho})K_1(\bar{\rho})K_2(\bar{\rho})+\bar{\rho} K_0(\bar{\rho})K_1(\bar{\rho})^2+\bar{\rho} K_1(\bar{\rho})^2K_2(\bar{\rho})\\
&&+\bar{\Lambda}\left[-\frac{\bar{\rho}}{2}K_0(\bar{\rho})^2K_1(\bar{\rho}+2\bar{\rho} K_1(\bar{\rho})^3+2K_0(\bar{\rho})K_1(\bar{\rho})^2+K_0(\bar{\rho})K_2(\bar{\rho})^2\right.\nonumber\\
&&-K_0(\bar{\rho})^2K_2(\bar{\rho})\left.+4K_1(\bar{\rho})^2K_2(\bar{\rho})+\frac{\bar{\rho} }{2}K_1(\bar{\rho})K_2(\bar{\rho})^2\right.\nonumber\\
&&\left.+\left(\frac{4}{\bar{\rho} }-2\bar{\rho}\right)K_0(\bar{\rho})K_1(\bar{\rho})K_2(\bar{\rho})\right]+\bar{\Lambda}^2\left[2K_0(\bar{\rho})^2K_1(\bar{\rho})+\frac{6}{\bar{\rho} }K_0(\bar{\rho})^2K_2(\bar{\rho})\right.\nonumber\\
&&\left.+\frac{2}{\bar{\rho}}K_0(\bar{\rho})K_2(\bar{\rho})^2+2K_1(\bar{\rho})K_2(\bar{\rho})^2\right],\nonumber
\end{subeqnarray}
\begin{equation}
B=\frac{1}{C_{\rm de}}\left\{-\bar{\rho} K_1(\bar{\rho})^2+\bar{\Lambda}\left[-\frac{5}{2}\bar{\rho} K_0(\bar{\rho})K_1(\bar{\rho})-\frac{\bar{\rho}}{2}K_1(\bar{\rho})K_2(\bar{\rho})\right]-2\bar{\Lambda}^2\bar{\rho} K_1(\bar{\rho})^2\right\},
\end{equation}
and
\begin{align}
\begin{split}
A=&\frac{1}{C_{\rm de}}\left\{2K_1(\bar{\rho})K_2(\bar{\rho})+\bar{\Lambda}\left[2K_1(\bar{\rho})^2+3K_0(\bar{\rho})K_2(\bar{\rho})+\frac{4}{\bar{\rho}}K_1(\bar{\rho})K_2(\bar{\rho})+K_2(\bar{\rho})^2\right]\right.\\
&\left.+\bar{\Lambda}^2\left[2K_0(\bar{\rho})K_1(\bar{\rho})+\frac{6}{\bar{\rho}}K_0(\bar{\rho})K_2(\bar{\rho})+2K_1(\bar{\rho})K_2(\bar{\rho})+\frac{2}{\bar{\rho}}K_2(\bar{\rho})^2\right]\right\}.
\end{split}
\end{align}
According to \eqref{eq:bc2_wc_3}, 
the time-averaged swimming speed $w^{(1)}$ is zero, and as expected we thus  need to consider the problem at order $\epsilon^2$. 

Following the slip boundary conditions, the $z$-component of the flow at order two, $w^{(2)}$,  
 satisfies on the boundary 
\begin{align}\label{eq:bc2ord2_wc3}
\begin{split}
\left(w^{(2)}+ w_q'\cos^2\theta\sin^2 s+\frac{1}{r}w_q\sin^2\theta\sin^2 s\right)\bigg|_{r=\bar{\rho}}=&\bar{\Lambda}\biggl\{\frac{\partial w^{(2)}}{\partial r}+\frac{\partial u^{(2)}}{\partial z}\\
&+\sin^2 s\biggl[(w_q''-u_q')\cos^2\theta\\
&-\frac{1}{r}\left(\frac{w_q}{r}-w_q'+u_q+v_q\right)\sin^2\theta\biggr]\\
&+\cos^2 s\biggl[u_q'\cos^2\theta+\frac{u_q+v_q}{r}\sin^2\theta\biggr]\biggr\}\bigg|_{r=\bar{\rho}}.
\end{split}
\end{align}
Averaging this equation in time and along the azimuthal direction, we obtain explicitly the swimming velocity,  $U^{(2)}$, as 
\begin{align}\label{finalslip}
\begin{split}
4U^{(2)}&=w_q'(\bar{\rho})+\frac{w_q(\bar{\rho})}{\bar{\rho}}-\bar{\Lambda}\left[w_q''-\frac{1}{\bar{\rho}}\left(\frac{w_q}{\bar{\rho}}-w_q'\right)\right]\bigg|_{r=\bar{\rho}}\\
&=\left[\frac{2}{\bar{\rho}}K_1(\bar{\rho})-K_2(\bar{\rho})-\bar{\Lambda} K_1(\bar{\rho})\right]B+\left[\frac{2}{\bar{\rho}}K_1(\bar{\rho})-K_2(\bar{\rho})-\bar{\Lambda} K_1(\bar{\rho})\right]C\\
&+\left\{2K_0(\bar{\rho})-\bar{\rho} K_1(\bar{\rho})-\frac{2}{\bar{\rho}}K_1(\bar{\rho})+K_2(\bar{\rho})+\bar{\Lambda}[5K_1(\bar{\rho})-\bar{\rho} K_2(\bar{\rho})]\right\}A.
\end{split}
\end{align}

The result in Eq.~(\ref{finalslip}) can also be evaluated in the no-slip case   by simply setting $\bar{\Lambda} = 0$, and we recover Taylor's result, namely
\begin{equation}\label{Taylorfilament}
U^{(2)}_{\rm no-slip}=\frac{\bar{\rho} K_1(\bar{\rho})^2K_2(\bar{\rho})-\bar{\rho} K_0(\bar{\rho})K_2(\bar{\rho})^2}{-2\bar{\rho} K_0(\bar{\rho})K_1(\bar{\rho})^2+4\bar{\rho} K_0(\bar{\rho})^2K_2(\bar{\rho})-4K_0(\bar{\rho})K_1(\bar{\rho})K_2(\bar{\rho})-2\bar{\rho} K_1(\bar{\rho})^2K_2(\bar{\rho})}\cdot
\end{equation}

\begin{figure}[t]
\includegraphics[width=\textwidth]{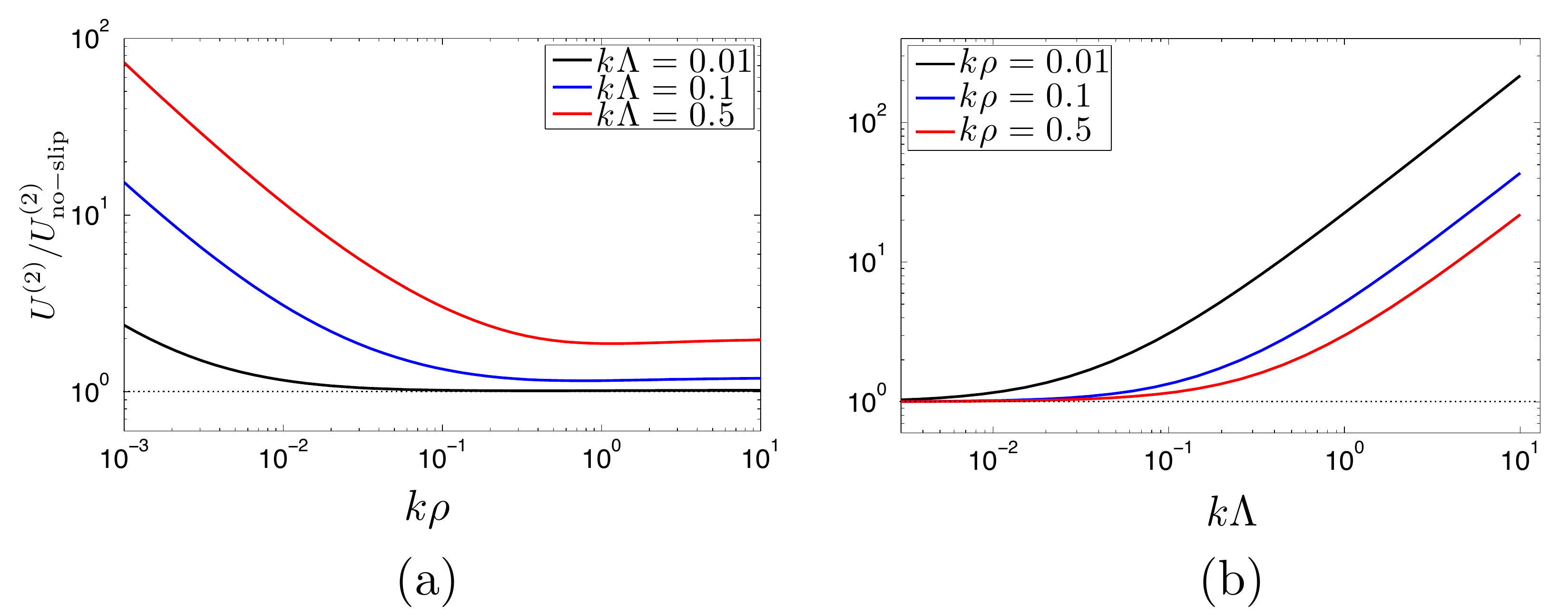}
\caption{{\color{black}{(Color online)}} Ratio of the swimming velocity in the case of  slip to the no-slip value, $U^{(2)}/U_{\rm no-slip}^{(2)}$. (a): Dependence on the dimensionless filament radius, $k\rho $, for three wave numbers $(k\Lambda =  0.01$, $0.1$ and $0.5$); (b): Dependence on the 
dimensionless wave number (with {\color{black}{$k\rho = 0.01$, 0.1 and 0.5}}).}\label{ratio_wc}
\end{figure}

The ratio between the swimming speed in the slip case to that in the no-slip situation,  $U^{(2)}/U_{\rm no-slip}^{(2)}$,  is plotted in   Fig.~\ref{ratio_wc} as a function of the dimensionless filament radius (Fig.~\ref{ratio_wc}a) and the dimensionless wave number  (Fig.~\ref{ratio_wc}b). As in the two-dimensional situation, the presence of slip is seen to always lead to faster swimming than in the no-slip case, and here the effect  can be potentially very large (the applicability of these results to recent experiments is  discussed in \S\ref{sec:discussion}). In Fig.~\ref{ratio_wc} we see that the swimming speed increases monotonically when either  the length scale of the swimmer cross section $(\rho)$ or the typical length scale of the waving motion, $k^{-1}$  {\color{black}becomes} smaller than  the slip length. In the opposite limit, the   no-slip result, Eq.~(\ref{Taylorfilament}),   is recovered when all length scales are much larger than $\Lambda$. 

We  further note that it is possible to  compute the swimming speed for small slip length as a power expansion in $k \Lambda$ (i.e.~$\bar{\Lambda}$),  $U^{(2)}=U^{(2)}_0+(k \Lambda )U^{(2)}_1+\dots$, with $U^{(2)}_0$ being  the no-slip swimming speed. When $k \rho$ (i.e.~$\bar{\rho}$) increases to infinity, {\color{black}the radius of the cylinder becomes much larger than any other length  scale,}  and we recover  $U^{(2)}_1/U^{(2)}_0=2$, leading to ${U^{(2)}}/{U_{\rm no-slip}^{(2)}}=1+2 k \Lambda$, 
which as expected agrees with the  results for the two-dimensional sheet.  

{\color{black}  We conclude by pointing out that although the calculation above was carried out in the case of planar waving deformation, similar algebra would govern swimming by propagating helical waves \cite{chwang71}, and in that case $\rho$ would be the radius of the helical flagellum, or that of the bundle of flagella in the case of  bacteria with multiple flagellar filaments such as \emph{E.~coli}.}

\section{Swimming in a two-fluid domain}
\label{sec:two-fluid}
In the previous section, we modeled the influence of phase separation as due to a finite apparent slip length, and showed that it leads to a systematic enhancement of the swimming speeds. In order to provide an alternative  microscopic physical picture, we instead consider in this section a second model where we include explicitly the presence of a low-viscosity layer near the surface of the swimmer, and we apply it to the two canonical swimmers (waving sheet and filament) considered in the previous section. 
\begin{figure}[b]
\centering
\includegraphics[width=4.6in]{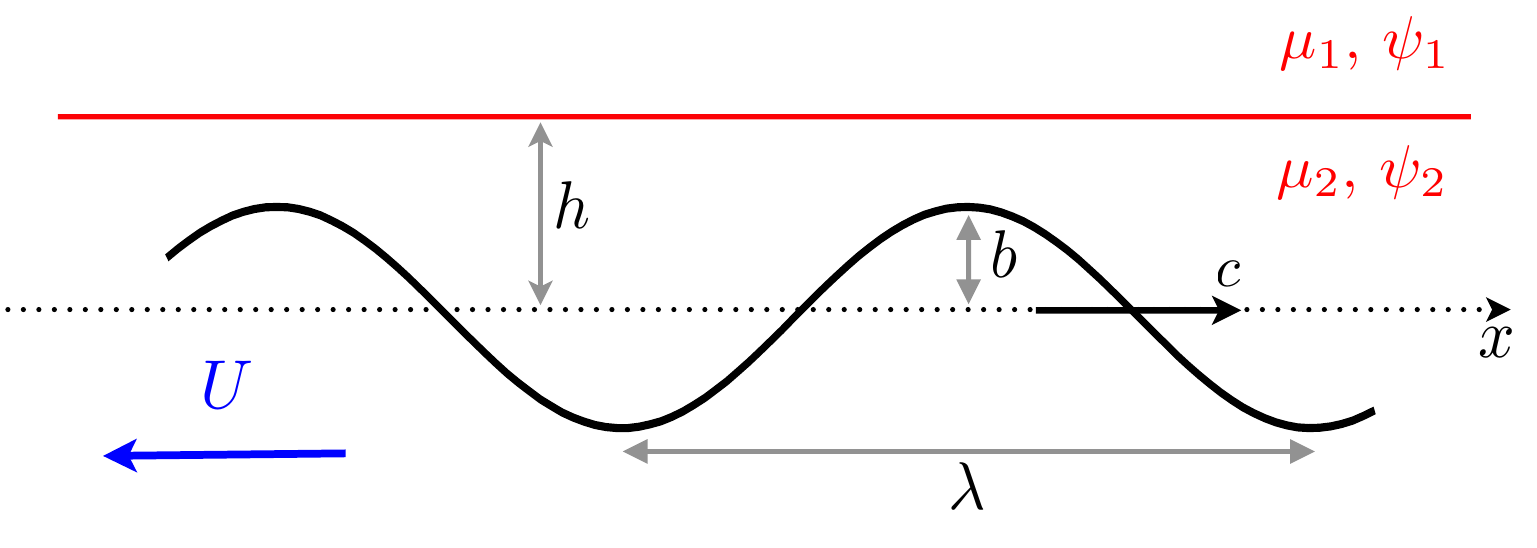}
\caption{{\color{black}{(Color online)}} Geometry of two-dimensional waving sheet swimming in a two-fluid domain. The bulk fluid has viscosity $\mu_1$ and the thin layer near the swimmer, of mean thickness $h$,  has viscosity $\mu_2 < \mu_1$. The two  streamfunctions in the fluids are denoted  $\psi_1$ and $\psi_2$. The interface between the two fluids is assumed to remain flat.}\label{geo_2s}
\end{figure}

\subsection{Two dimensional waving sheet}
\label{sec:twodfluid:A}
We first consider the swimming sheet setup shown in Fig.~\ref{geo_2s}. 
 The fluid is composed of two domains: the bulk fluid has viscosity $\mu_1$ while the thin layer near the swimmer has a smaller viscosity  $\mu_2$ ($\mu_1>\mu_2$). The average distance between the sheet and the fluid-fluid interface, which is the  thickness of the low-viscosity layer, is denoted $h$ and assumed to remain constant ({\color{black}this assumption is discussed in \S\ref{sec:discussion}}). All other notation  are similar to the ones in \S\ref{sec:sheet-slip}.  Following the same  nondimentionalization, we now have the no-slip boundary conditions on the sheet
\begin{equation}
\displaystyle\frac{\partial\psi_2}{\partial y}(x, \epsilon \sin\xi)=0,\,\, \frac{\partial\psi_2}{\partial x}(x, \epsilon \sin\xi)=-\epsilon \cos\xi,\\
\end{equation}
while at infinity we have the unknown swimming speed
\begin{equation}
\displaystyle \frac{\partial\psi_1}{\partial y}(x, \infty)=-U, \,\,\frac{\partial\psi_1}{\partial x}(x, \infty)=0.
\end{equation}

At the flat interface between the two fluids, we have continuity of velocities 
\begin{equation}
\displaystyle\frac{\partial\psi_1}{\partial y}(x, \bar{h})=\frac{\partial\psi_2}{\partial y}(x, \bar{h}),\,\, \frac{\partial\psi_1}{\partial x}(x, \bar{h})=\frac{\partial\psi_2}{\partial x}(x, \bar{h})=0,
\end{equation}
and here again we keep the ``bar'' notation for  the thickness $\bar h\equiv kh $. Together with continuity of tangential stresses, which is written as
\begin{equation}
\displaystyle\left(\frac{\partial^2\psi_1}{\partial y^2}-\frac{\partial^2\psi_1}{\partial x^2}\right)\bigg|_{(x, \bar{h})}=\beta\left(\frac{\partial^2\psi_2}{\partial y^2}-\frac{\partial^2\psi_2}{\partial x^2}\right)\bigg|_{(x, \bar{h})},
\end{equation}
where $\beta$ denotes the ratio of viscosities, $\beta= \mu_2/\mu_1<1$. 

As in \S\ref{sec:sheet-slip} we solve the problem as a perturbation expansion in $\epsilon$. The general periodic solution for to the biharmonic equation  which vanishes at infinity is obtained by separation of variables as
\begin{equation}
V_n=(a_ny+b_n)\sin n\xi e^{-ny}+(c_n y+d_n)\cos n\xi e^{-ny},
\end{equation}
which we use to expand  $\psi_1$  as
\begin{equation}
\psi_1=\epsilon\psi_1^{(1)}+\epsilon^2\psi_2^{(2)}+\dots
\end{equation}
with 
\begin{align}
\psi_1^{(m)}=-U^{(m)}y + V_1^{(m)}+\dots+V_m^{(m)}.
\end{align}
Similarly, in the second domain we expand $\psi_2$ as
\begin{align}
\psi_2&=\epsilon\psi_2^{(1)}+\epsilon^2\psi_2^{(2)}+\dots\\
\psi_2^{(m)}&=W_1^{(m)}+\cdots+W_m^{(m)}+\chi^{(m)}y^2+\eta^{(m)}y,
\end{align}
where $W_n$ is the general  periodic solution for to the biharmonic equation  in a finite domain  obtained by separation of variables 
\begin{align}
\begin{split}
W_n&=[(A_ny+B_n)\sin n\xi+(C_ny+D_n)\cos n\xi]\sinh ny\\
&+[(E_ny+F_n)\sin n\xi+(G_ny+H_n)\cos n\xi]\cosh ny.
\end{split}
\end{align}

Expanding the boundary conditions around $y=0$, we obtain at first order
\begin{subeqnarray}
\text{(on sheet)}&\displaystyle\frac{\partial\psi_2\ord1}{\partial y}(x, 0)=0 , \frac{\partial\psi_2\ord1}{\partial x}(x, 0)=-\cos\xi\\
\text{(on interface)}&\displaystyle\frac{\partial\psi_1\ord1}{\partial y}(x, \bar{h})=\frac{\partial\psi_2\ord1}{\partial y}(x, \bar{h}),\,\, \frac{\partial\psi_1\ord1}{\partial x}(x, \bar{h})=\frac{\partial\psi_2\ord1}{\partial x}(x, \bar{h})=0\\
&\displaystyle\left(\frac{\partial^2\psi_1\ord1}{\partial y^2}-\frac{\partial^2\psi_1\ord1}{\partial x^2}\right)\bigg|_{(x, \bar{h})}=\beta\left(\frac{\partial^2\psi_2\ord1}{\partial y^2}-\frac{\partial^2\psi_2\ord1}{\partial x^2}\right)\bigg|_{(x, \bar{h})}\\
\text{(at infinity)}&\displaystyle \frac{\partial\psi_1\ord1}{\partial y}(x, \infty)=-U\ord1, \,\,\frac{\partial\psi_1\ord1}{\partial x}(x, \infty)=0.
\end{subeqnarray}
Substituting these boundary conditions into the general solution we obtain the coefficients at first order
\begin{subeqnarray}
U\ord1&=&\chi^{(1)}=\eta^{(1)}=c_1^{(1)}=d_1^{(1)}=C_1^{(1)}=D_1^{(1)}{\color{black}{=0}},\\
a_1\ord1&=&\frac{\beta \bar{h}e^{\bar{h}}\sinh \bar{h}}{\sinh^2\bar{h}-\bar{h}^2+\beta(\sinh \bar{h}\cosh \bar{h}-\bar{h})},\\
b_1\ord1&=&\frac{-\beta \bar{h}^2e^{\bar{h}}\sinh \bar{h}}{\sinh^2\bar{h}-\bar{h}^2+\beta(\sinh \bar{h}\cosh \bar{h}-\bar{h})},\\
A_1\ord1&=&\frac{\sinh^2\bar{h}+\beta(\sinh \bar{h}\cosh \bar{h})}{\sinh^2\bar{h}-\bar{h}^2+\beta(\sinh \bar{h}\cosh \bar{h}-\bar{h})},\\
B_1\ord1&=&\frac{\sinh \bar{h}\cosh \bar{h}+\bar{h}+\beta\cosh^2\bar{h}}{\sinh^2\bar{h}-\bar{h}^2+\beta(\sinh \bar{h}\cosh \bar{h}-\bar{h})}\cdot
\end{subeqnarray}

At second order, the  boundary conditions become
\begin{subeqnarray}
\text{(on sheet)}&\displaystyle\frac{\partial\psi_2\ords}{\partial y}(x, 0)+\sin\xi\frac{\partial^2\psi_2\ord1}{\partial y^2}(x,0)=0,\\
\text{(on interface)}&\displaystyle\frac{\partial\psi_1\ords}{\partial y}(x, \bar{h})=\frac{\partial\psi_2\ords}{\partial y}(x, \bar{h}),\\
&\displaystyle\left(\frac{\partial^2\psi_1\ords}{\partial y^2}-\frac{\partial^2\psi_1\ords}{\partial x^2}\right)\bigg|_{(x, \bar{h})}=\beta\left(\frac{\partial^2\psi_2\ords}{\partial y^2}-\frac{\partial^2\psi_2\ords}{\partial x^2}\right)\bigg|_{(x, \bar{h})},
\end{subeqnarray}
leading to the second-order swimming speed as
\begin{equation}
U\ords=-\eta\ords=A_1\ord1-\frac{1}{2},
\end{equation}
and therefore
\begin{equation}\label{final_2_sheet}
U\ords=\frac{1}{2}+\frac{\bar{h}^2+\beta \bar{h}}{\sinh^2\bar{h}-\bar{h}^2+\beta(\sinh \bar{h}\cosh \bar{h}-\bar{h})}\cdot
\end{equation}

In Eq.~(\ref{final_2_sheet}), the first term is the one-fluid classical result of Taylor ($U\ords_{\infty}=1/2$, recovered when $\bar{h}\to\infty$) and 
the second fraction is always  positive  since $\bar{h} > 0$. As a consequence the swimming speed for a waving sheet in a two-fluid domain is always faster than in a homogeneous Newtonian fluid. 

\begin{figure}[t]
\includegraphics[width=3.4in]{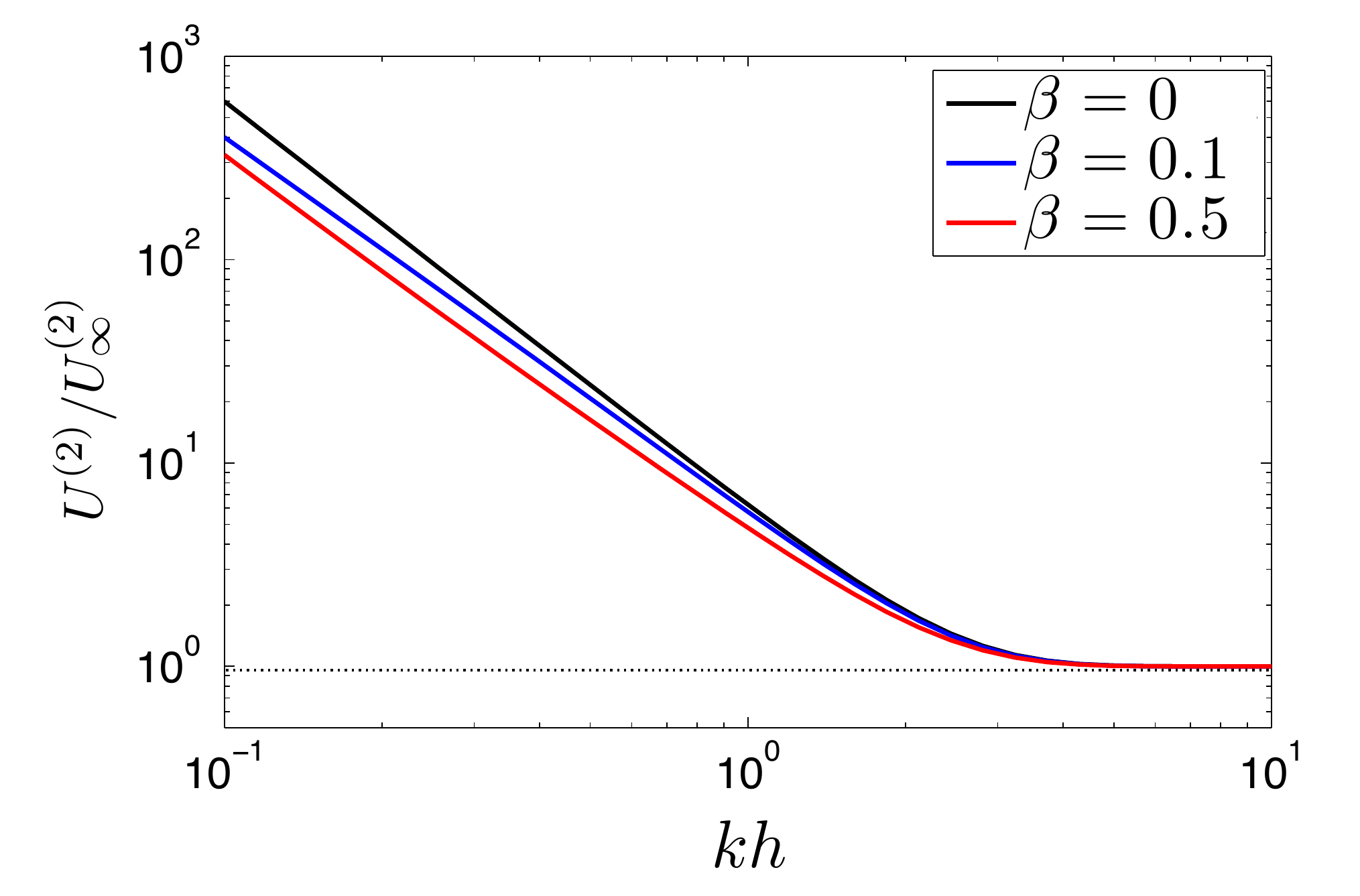}
\caption{{\color{black}{(Color online)}} Ratio between the swimming velocity in the two-fluid domain, ${U\ords}$, and the velocity in the presence of a single fluid, ${U\ords_{\infty}}$, as a function of the dimensionless distance between the swimmer and the interface, $kh$, for three values of the viscosity ratio: $\beta=0$, 0.1 and 0.5.}\label{2ws}
\end{figure}

We display in  Fig.~\ref{2ws} the ratio  $U\ords/U\ords_{\infty}$ as a function of the dimensionless distance to the interface, $kh$ (i.e.~$\bar{h}$). 
We observe that the increase of the swimming speed can become very large when the thickness of the low-shear layer is smaller than the wavelength of the swimmer. We also see that the overall conclusions and speed ratios are rather insensitive to the exact value of the viscosity ratio, $\beta$.


\subsection{Three dimensional waving filament}
\begin{figure}[t]
\centering
\includegraphics[width=2.5in]{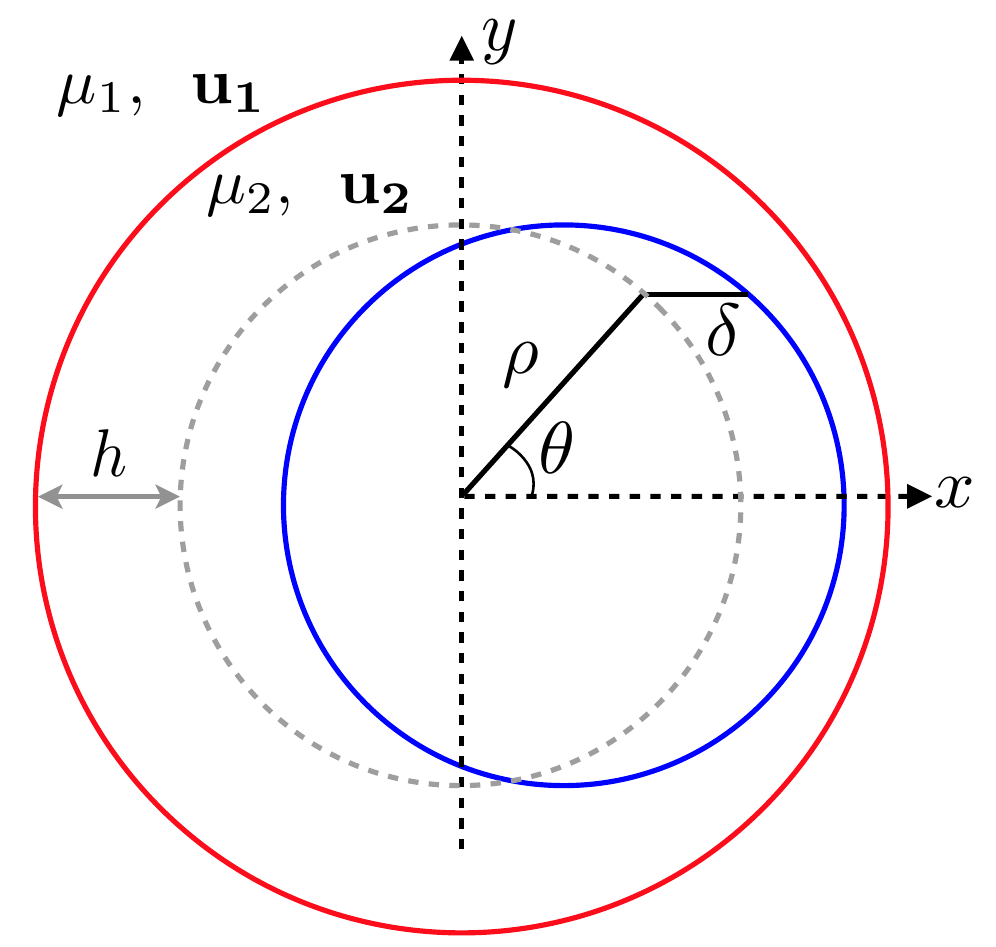}
\caption{{\color{black}{(Color online)}} Waving motion of amplitude $\delta$ in the  circular cross-section  of a three-dimensional waving filament of radius $\rho$ in a two-fluid domain. The solid blue line {\color{black}{(inside)}} indicates the instantaneous position of the filament,  the  dashed gray  line the average location of the cross section, while the solid red  line {\color{black}{(outside)}} shows the interface between the thin low-viscosity layer (mean thickness $h$; dynamic viscosity $\mu_2$) and the bulk fluid (viscosity $\mu_1 > \mu_2$).
}\label{geo_2wc}
\end{figure}

In this final section, we  extend the two-fluid scenario to the case of  three-dimensional waving filaments. In this case, the geometry of the cross-section,   shown in Fig.~\ref{geo_2wc}, is analogous to the one addressed in 
\S\ref{sec:3d:1} with the added ingredient that we now have two fluids. The thin, low-viscosity layer, has mean  thickness $h$ and dynamic viscosity $\mu_2$ while the bulk has viscosity $\mu_1 > \mu_2$. All other notation are similar to the ones used in \S\ref{sec:3d:1}.

On the surface of the filament (dimensionless form), $\r_s=(\delta+\bar{\rho}\cos\theta)\x+\bar{\rho}\sin\theta\y+z\z$, we have the distribution of surface velocities $\u_2(\delta+\bar{\rho}\cos\theta,\theta,z)=\epsilon\cos s\x$, 
which can be written in polar coordinates as
\begin{equation}
u_2(\delta+\bar{\rho}\cos\theta,\theta,z)=\epsilon\cos\theta\cos s, \quad v_2(\delta+\bar{\rho}\cos\theta,\theta,z)={\color{black} -\epsilon \sin\theta\cos s}.
\end{equation}
Expanding the velocity components  around the averaged position of the surface Eq.~(\ref{eq:ex_ap}), we obtain at first order
\begin{equation}
u_2^{(1)}(\bar{\rho},\theta,z)=\cos\theta\cos s, \quad v_2^{(1)}(\bar{\rho},\theta,z)=-\sin\theta\cos s, \quad w_2^{(1)}(\bar{\rho},\theta,z)=0.
\end{equation}

We assume that the interface $r=\bar{h}+\bar{\rho}\equiv l$ undergoes no radial motion and apply continuity of  the tangential components of velocities  and traction leading to the conditions
\begin{subeqnarray}\label{eq:bc2f_wc}
u_1^{(1)}(l,\theta,z)=u_2^{(1)}(l,\theta,z)=0, \\
v_1^{(1)}(l,\theta,z)=v_2^{(1)}(l,\theta,z), \quad w_1^{(1)}(l,\theta,z)=w_2^{(1)}(l,\theta,z),\\
\left(\frac{\partial w_1^{(1)}}{\partial r}+\frac{\partial u_1^{(1)}}{\partial z}\right)\bigg|_{r=l}=\beta\left(\frac{\partial w_2^{(1)}}{\partial r}+\frac{\partial u_2^{(1)}}{\partial z}\right)\bigg|_{r=l},\\
\left[r\frac{\partial(v_1^{(1)}/r)}{\partial r}+\frac{1}{r}\frac{\partial u_1^{(1)}}{\partial\theta}\right]\bigg|_{r=l}=\beta\left[r\frac{\partial(v_2^{(1)}/r)}{\partial r}+\frac{1}{r}\frac{\partial u_2^{(1)}}{\partial\theta}\right]\bigg|_{r=l},
\end{subeqnarray}
where, as in \S\ref{sec:twodfluid:A}, $\beta$ denotes the ratio of viscosity, $\beta = \mu_2/\mu_1 < 1$. 

Using separation of variables, we write
\begin{subeqnarray}\label{sep}
u_1^{(1)}=u_{1q}(r)\cos\theta\cos s,\quad v_1^{(1)}=v_{1q}(r)\sin\theta\cos s, \quad w_1^{(1)}=w_{1q}(r)\cos\theta\sin s,\\
u_2^{(1)}=u_{2q}(r)\cos\theta\cos s, \quad v_2^{(1)}=v_{2q}(r)\sin\theta\cos s, \quad w_2^{(1)}=w_{2q}(r)\cos\theta\sin s,
\end{subeqnarray}
and substituting Eq.~(\ref{sep}) into the Eq.~(\ref{eq:bc2f_wc}), the boundary conditions become
\begin{subeqnarray}\label{eq:bc2f2_wc}
u_{2q}(\bar{\rho})=1,\quad  v_{2q}(\bar{\rho})=-1,\quad  w_{2q}(\bar{\rho})=0,\\
u_{1q}(l)=u_{2q}(l)=0, \quad v_{1q}(l)=v_{2q}(l), \quad  w_{1q}(l)=w_{2q}(l),\\
w_{1q}'(l)-u_{1q}(l)=\beta[w_{2q}'(l)-u_{2q}(l)],\\
{\color{black}v_{1q}'(l)}-\frac{v_{1q}(l)}{l}-\frac{u_{1q}(l)}{l}=\beta\left[v_{2q}'(l)-\frac{v_{2q}(l)}{l}-\frac{u_{2q}(l)}{l}\right].
\end{subeqnarray}
{\color{black}We then using the general solution in Eq.~(\ref{Bessels}), with coefficients $A_1$, $B_1$, $C_1$ for fluid \#1, ensuring the correct decay in the far field,  and coefficients $A_2$, $B_2$, $C_2$, $D$, $E$, $F$ for fluid \#2.}

Substituting this general solution into the boundary conditions in Eq.~(\ref{eq:bc2f2_wc}), we obtain a linear system with 9 coefficients,
\begin{subeqnarray}\label{horrible}
\bar{\rho} K_1(\bar{\rho})A_2+K_2(\bar{\rho})B_2+K_0(\bar{\rho})C_2+\bar{\rho} I_1(\bar{\rho})D+I_2(\bar{\rho})E+I_0(\bar{\rho})F=1,\,\, \\
K_2(\bar{\rho})B_2-K_0(\bar{\rho})C_2+I_2(\bar{\rho})E-I_0(\bar{\rho})F=-1,\,\, \\
\left[\bar{\rho} K_0(\bar{\rho})-K_1(\bar{\rho})\right]A_2+K_1(\bar{\rho})B_2+K_1(\bar{\rho})C_2-\left[\bar{\rho} I_0(\bar{\rho})+I_1(\bar{\rho})\right]D\notag\,\, \\
-I_1(\bar{\rho})E-I_1(\bar{\rho})F=0,\,\, \\
lK_1(l)A_1+K_2(l)B_1+K_0(l)C_1=0,\,\, \\
lK_1(l)A_2+K_2(l)B_2+K_0(l)C_2+lI_1(l)D+I_2(l)E+I_0(l)F=0,\,\, \\
K_2(l)B_1-K_0(l)C_1-K_2(l)B_2+K_0(l)C_2-I_2(l)E+I_0(l)F=0,\,\, \\
\left[lK_0(l)-K_1(l)\right]A_1+K_1(l)B_1+K_1(l)C_1-[lK_0(l)-K_1(l)]A_2-K_1(l)B_2-K_1(l)C_2\notag\,\, \\
+[lI_0(l)+I_1(l)]D+I_1(l)E+I_1(l)F=0,\,\,\\
\left[2K_0(l)-2lK_1(l)+\frac{K_1(l)}{l}\right]A_1+\left[\frac{K_1(l)}{l}-2K_2(l)\right]B_1-\left[2K_0(l)-\frac{K_1(l)}{l}\right]C_1\notag \,\, \\
-\beta\left[2K_2(l)-2lK_1(l)+\frac{K_1(l)}{l}\right]A_2-\beta\left[\frac{K_1(l)}{l}-2K_2(l)\right]+\beta\left[2K_0(l)+\frac{K_1(l)}{l}\right]C_2\notag \,\, \\
+\beta\left[2I_0(l)+2lI_1(l)-\frac{I_1(l)}{l}\right]D+\beta\left[\frac{I_1(l)}{l}+2I_2(l)\right]E+\beta\left[2I_0(l)-\frac{I_1(l)}{l}\right]F=0,\,\, \\
-K_1(l)A_1-K_3(l)B_1+K_1(l)C_1+\beta[ K_1(l)A_2+K_3(l)B_2-K_1(l)C_2\notag\,\, \\
+I_1(l)D-I_3(l)E+I_1(l)F]=0,\,\, 
\end{subeqnarray}
which can be easily inverted numerically. 

The last step consists in moving to next order and computing the swimming speed. This is done similarly to the case with a finite slip length, and we apply 
Eq.~(\ref{eq:bc2ord2_wc3}) with $\Lambda=0$. The average of $w_2^{(2)}$ on the filament is the swimming speed,  $U^{(2)}$, and  we have
\begin{equation}
U^{(2)}=\frac{1}{4}\left[w_{2q}'(\bar{\rho})+\frac{w_{2q}(\bar{\rho})}{\bar{\rho}}\right],
\end{equation}
which can be evaluated as
\begin{align}
\begin{split}
U^{(2)}=&\frac{1}{4}\left\{-B_2K_0(\bar{\rho})-C_2K_0(\bar{\rho})+A_2[3K_0(\bar{\rho})-\bar{\rho} K_1(\bar{\rho})]\right.\\
&\left.-EI_0(\bar{\rho})-FI_0(\bar{\rho}) -D[3I_0(\bar{\rho})+\bar{\rho} I_1(\bar{\rho})]\right\}.
\end{split}
\end{align}

\begin{figure}[t]
\centering
\includegraphics[width=6in]{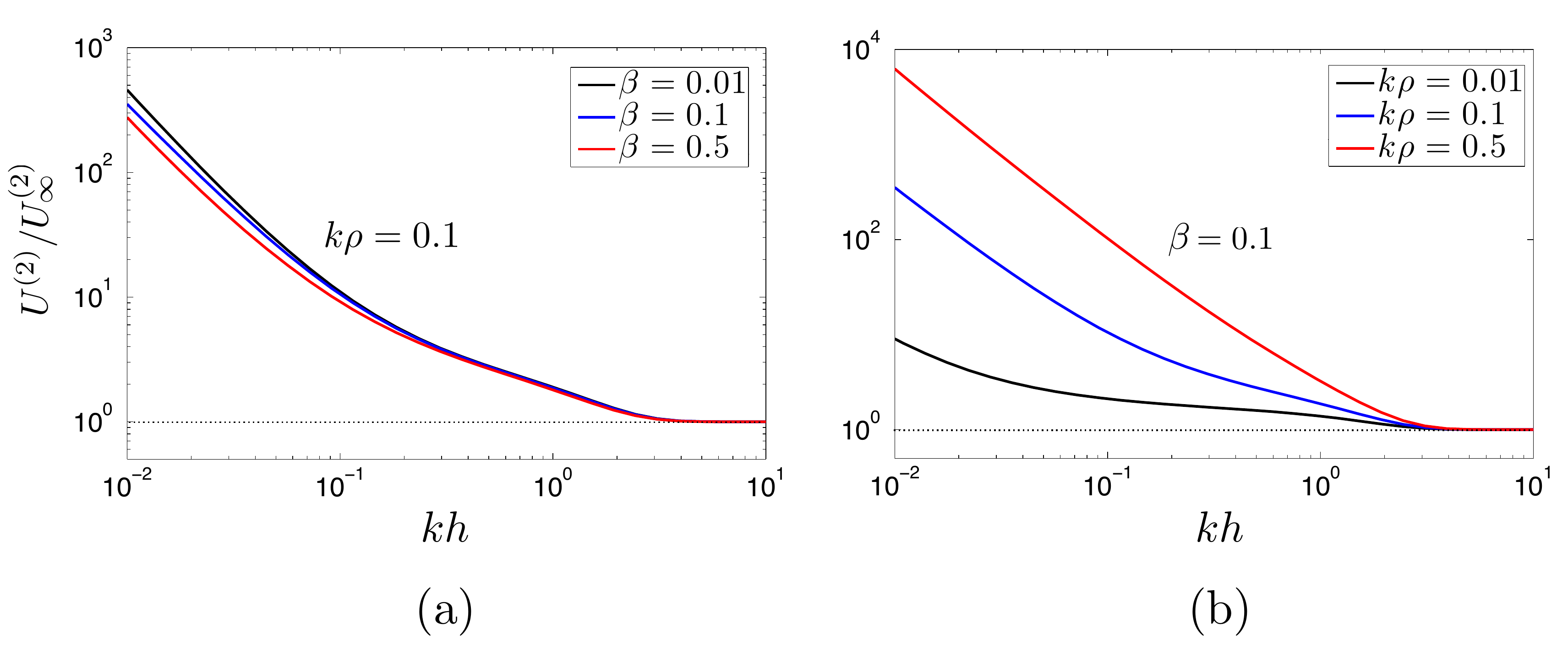}
\caption{{\color{black}{(Color online)}} Ratio between the swimming velocity in the two-fluid domain, $U\ords$, and that obtained in the case of a single Newtonian fluid, $U\ords_{\infty}$, as a function of the mean dimensionless  distance  between the cylindrical filament and the fluid interface, $kh$;  
(a): the dimensionless radius is fixed,  $k\rho =0.1$, and three values of  viscosity ratios are considered ($\beta=0.01$, $0.1$ and $0.5$); 
(b): fixed viscosity ratio ($\beta= 0.1$), and three different values of dimensionless radius ($k\rho =  0.01$, $0.1$ and $0.5$).}
\label{2wc}
\end{figure}  

In Fig.~\ref{2wc} we plot the ratio between the swimming speed of the filament in the two-fluid domain, $U\ords$, and that obtained in the case of a simple fluid, $U\ords_{\infty}$, for a range of values of the dimensionless radius ($k\rho$, i.e.~$\bar{\rho}$) and viscosity ratio ($\beta$). The results are reminiscent of the ones shown in Fig.~\ref{2ws} for the two-dimensional case. The swimming speed is always increased by the presence of a second fluid, potentially by order of magnitude when the wavelength of the swimmer and its radius are  large compared to the thickness of the low-viscosity layer.

\section{Discussion}  
\label{sec:discussion}

In this paper, we presented a physical mechanism for the locomotion enhancement of microscopic swimmers  in a complex fluid. The physical idea is that   phase-separation near the surface of the swimmer leads to the presence of a low viscosity layer which  promotes slip and decreases viscous friction. As a way to 
intuitively rationalize the results in our paper, we note -- as is well known -- that the locomotion in a fluid is governed by the ratio between  the drag coefficients for the motion relative to the fluid perpendicular to and along the surface of the swimmer \cite{Powers2009}.  The presence of a {\color{black} fluidic interface}, or of a finite slip length,  affects normal hydrodynamic forces only weakly but leads to a systematic decrease of tangential viscous forces, and hence should always lead to faster swimming, as observed here. In  a different context, but with some physical similarities, swimmers also always enhance their swimming speed in a network of polymer molecules \cite{Magariyama02} and in a porous medium~\cite{leshansky2009}. 

{\color{black}  Beyond the traditional geometrical assumptions made in our paper which are similar to  a number of classical studies (namely solving the swimming problems for small amplitude motion and perfectly sinusoidal waveforms), one severe restriction of our two-fluid approach is the assumption that the interface between the two fluids remains flat. This is akin to saying that the time scale of the waving motion is much faster than the time scale for the readjustment of the interface, which is a reasonable assumption only for large fluid viscosities. That flat interface then provides an effective confinement to the swimmer, which is known to  enhance locomotion \cite{Reynolds1965}. A more sophisticated physicochemical model including molecular details of the phase separation would be required to solve for the dynamics of the thin film and to 
 untangle the relative importance of viscosity difference vs.~confinement in the increase of the swimming speed.}

What are the quantitative predictions of our models? In two dimensions, we obtained that the speed of a two-dimensional infinite swimming sheet is increased by $1+2k\Lambda$ for the swimming with wave number $k$. Slip lengths  of polymer solutions, $\Lambda$,  have been measured in the  range  $0.1-10$~$\mu$m \cite{Mhetar98}. For a microswimmer with wavelength  $\lambda = 2\pi/k \approx 10$~$\mu$m \cite{Powers2009}, the swimming speed can then be increased by $O(10\%)$ up to by one order of magnitude.   In the three-dimensional case, we also obtained the increased speed shown in Fig.~\ref{ratio_wc}, which is consistent with the two-dimensional situation. Considering the same wavelength, and for a filament with radius $\rho = 150$~nm (so with dimensionless radius $k\rho\approx 0.1$), the predicted  enhancement ranges from  $O(30\%)$ up to by  forty times of the speed in the Newtonian fluid.     

In our second model, we used a two-fluids domain to describe wall depletion. In both two- and three-dimensions, we saw that when the wall depletion layer is very thin, the enhancement can be very large.  A recent experiment by Gagnon, Shen and Arratia~\cite{GagnonShenEPL13} considered the locomotion of the nematode  \emph{C. elegans} in  concentrated polymer solutions and showed that the swimming  speed can be 
 increased significantly,  by up to $65\%$. Is our model consistent with this result? The thickness of the low-viscosity layer in concentrated polymer solutions is a complex function of mean diameter of the particles and the concentration but can be  estimated using the empirical formula   \cite{Kalyon05}
\begin{equation}
\frac{h}{D_p}=1-\frac{\phi}{\phi_m},
\end{equation}
where $D_p$ is the mean diameter of the particles, $\phi$ the particle volume fraction,  and $\phi_m$  the maximum packing fraction. The diameter of the particle (Xanthan gum in the experiment of Ref.~\cite{GagnonShenEPL13}) is about $200$~$\mu$m. For semi-concentrated solution, $\phi/\phi_m$ is 0.4, while for a concentrated solution the value is 0.8, so the slip layer thickness is around $120$~$\mu$m and $40$~$\mu$m respectively. The wavelength of the swimming worm  is $2\pi/k \approx 1$~mm, and the dimensionless diameter $k\rho$ is about 0.1. When the ratio of the viscosities $\beta$ changes from 0.01 to 0.5,  the enhancement predicted by our model is about $90-98\%$, which is less than a factor of two away from the experimental results, and indicates that our simplified approach captures the essential physics of the swimming enhancement.

\section*{Acknowledgements}

{\color{black} We  thank an anonymous referee for helpful insight.}
 This work was funded in part by the European Union through a Marie Curie grant CIG (E.L.) and by the Cambridge Commonwealth Trust and the Cambridge Overseas Trust (Y. M.). 


\end{document}